# Walking droplets, swimming microbes: on memory in physics and life

Albert Libchaber[1,*] and Tsvi Tlusty[2,3,*]

[1]Center for Studies in Physics and Biology, Rockefeller University, New York, NY 10021;
[2]Center for Soft and Living Matter, Institute for Basic Science, Ulsan 44919, Korea;
[3]Department of Physics, Ulsan National Institute of Science and Technology, Ulsan 44919, Korea.
*albert.libchaber@rockefeller.edu; tsvitlusty@gmail.com

*Abstract.* Whirling and swerving, a bacterium is swimming in a test tube, foraging for food. On the surface of a vibrating bath, a droplet starts walking. A certain similarity, but mostly dissimilarity, between the physical memory that emerges in Couder's droplet experiments and the biological memory of the bacterium is noted. It serves as a starting point for a short perspective and speculation on the multilevel, loopy memory of living matter.

*"…cet albâtre translucide de nos souvenirs…"* (A la recherche du temps perdu, M. Proust [1])

*"Consider what takes place when one is listening to music. At a given moment a certain note is being played but a number of the previous notes are still "reverberating" in consciousness. Close attention will show that it is the simultaneous presence and activity of all these reverberations that is responsible for the direct and immediately felt sense of movement, flow and continuity."*
                                                           (Wholeness and the Implicate Order, D. Bohm [2])

***Prologue.*** One of us (AL) has been interacting with Yves Couder on a permanent basis for the last fifty years. Lately, on various meetings, including the last one in his house just before his death, his mind was totally focused on the problem of memory, a fascinating observation in his extraordinary droplet experiment. He often asked me questions about biological memory that fascinated him. This paper is a response to his demand. We examine the bacterium *E.coli* as an example of biological memory and discuss how its multitude of memories are activated, depending on the type of stress. This paper is a tribute to Yves' creativity by both of us.

***On droplets and microbes.*** The most striking feature of Couder's droplet is the emergence of memory in a simple physical system. The common mechanical system is a slave to the moment: the state of the system at the present moment determines how it is going to change in the next one. Be it Newton's second law or

Schrödinger's equation, the dynamics is *local* in time, and how the space-time trajectory wiggled in in the past is irrelevant for its future. But Yves' droplet takes time quite differently (Figure 1 top, adapted from [3]). Coupling of their motion to the vibration of the bath awards them with a physical memory: the motion of a droplet is not governed merely by its instantaneous state but rather takes into account also its positions during previous collisions with the bath surface [3, 4]. This memory will gradually fade, but if friction is low, the decay is slow and the dynamics "remembers" far into the past, many recoils before the present one. There are other physical systems that exhibit various memory effects, such as hysteresis and shape memory, but these typically are heterogeneous many-body systems, such as glasses, or defects in alloys [5].

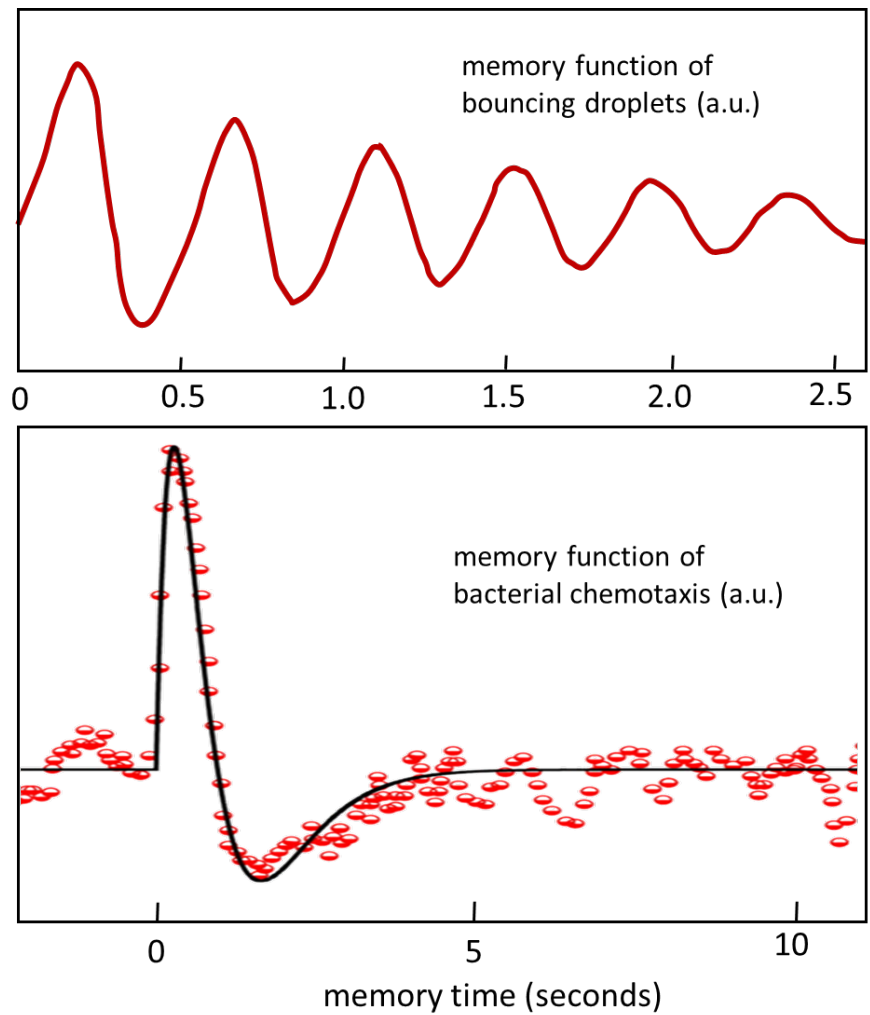

**Figure 1.** Memory functions of Couder's walking droplets (top) and of *E. coli* during chemotaxis (bottom).

In living matter, memory is the rule rather than the exception. Evolution is inherently *non-local* and necessitates memory at all scales of time and space. Consider an *E. coli* bacterium swimming in fluid, searching for food. To survive, the bacterium must optimize and adapt its nutrition to the changing environment. For example, in a non-uniform environment that is richer in nutrients in one direction, the bacterium will migrate uphill the nutrient gradient by tuning its "run-and-tumble" motion. This adaptive algorithm of directional flow called *chemotaxis* [6] is hardwired in the bacteria as a genetically encoded biochemical circuit. The bacterium is about a micron in length, so it is too small to sense spatial concertation gradients that change on much longer length scales. Instead, the bacterium measures the gradients indirectly, in time, using a biochemical memory circuit that adds up the concentration along its past trajectory $c(t)$ weighed by a memory function $m(\Delta t, \tau)$ whose time scale is $\tau$ (Figure 1 bottom, adapted from [7, 8]). An increase in the cumulative sum signals the bacterium that it is moving towards richer regions, and it will adapt its swimming accordingly by reducing the tumbling frequency by a fraction $R(t)$, following a linear response rule [7, 8]:

$$R(t) = \int^t ds\, c(s)\, m(t-s, \tau),$$

$$m(\Delta t, \tau) = \frac{e^{-\Delta t/\tau}}{\tau}\left[\frac{\Delta t}{\tau} - \frac{1}{2}\left(\frac{\Delta t}{\tau}\right)^2\right], \quad \Delta t \geq 0.$$

(1)

The biochemical memory function $m(\Delta t, \tau)$ in Eq. (1) fades exponentially with $\Delta t$ as it looks further back into the past, with a time scale $\tau$ similar to that of the droplet's physical memory.

One might be tempted to stretch this analogy, but the apparent similarity between the motion of bacteria and droplets is only at one timescale, of roughly ~1 second. Unlike the droplet, the bacterium has a multitude of memories, all interconnected and operating at time scales that stretch from the molecular scale of milliseconds to the evolutionary scales of billions of years. This paper briefly surveys this cascade of biological clocks and memories in *E. coli*, which appears to be a hallmark of living matter. This is followed by tentative thoughts and speculation, in terms of a minimal toy model, on how evolution drives the cascade's emergence and its explosion over many orders of magnitude.

*The memory cascade of bacteria*. We start with a quick tour of the memory cascade of bacteria, from the fastest to the slowest time scales (Figure 2). We use the term 'cascade` also to invoke the analogy to the cascade of energy and time scales that occurs in a turbulent flow. In turbulence, energy is transferred along a hierarchy of flow scales, spanning from large-scale eddies down to microscales governed by viscous dissipation [9]. Energy transfer is bi-directional, from large to small scales in the *direct cascade* and small to large scales in the *inverse cascade*. Likewise, the bidirectional signaling through the bacterium's memory cascade is essential to its survival in evolution.

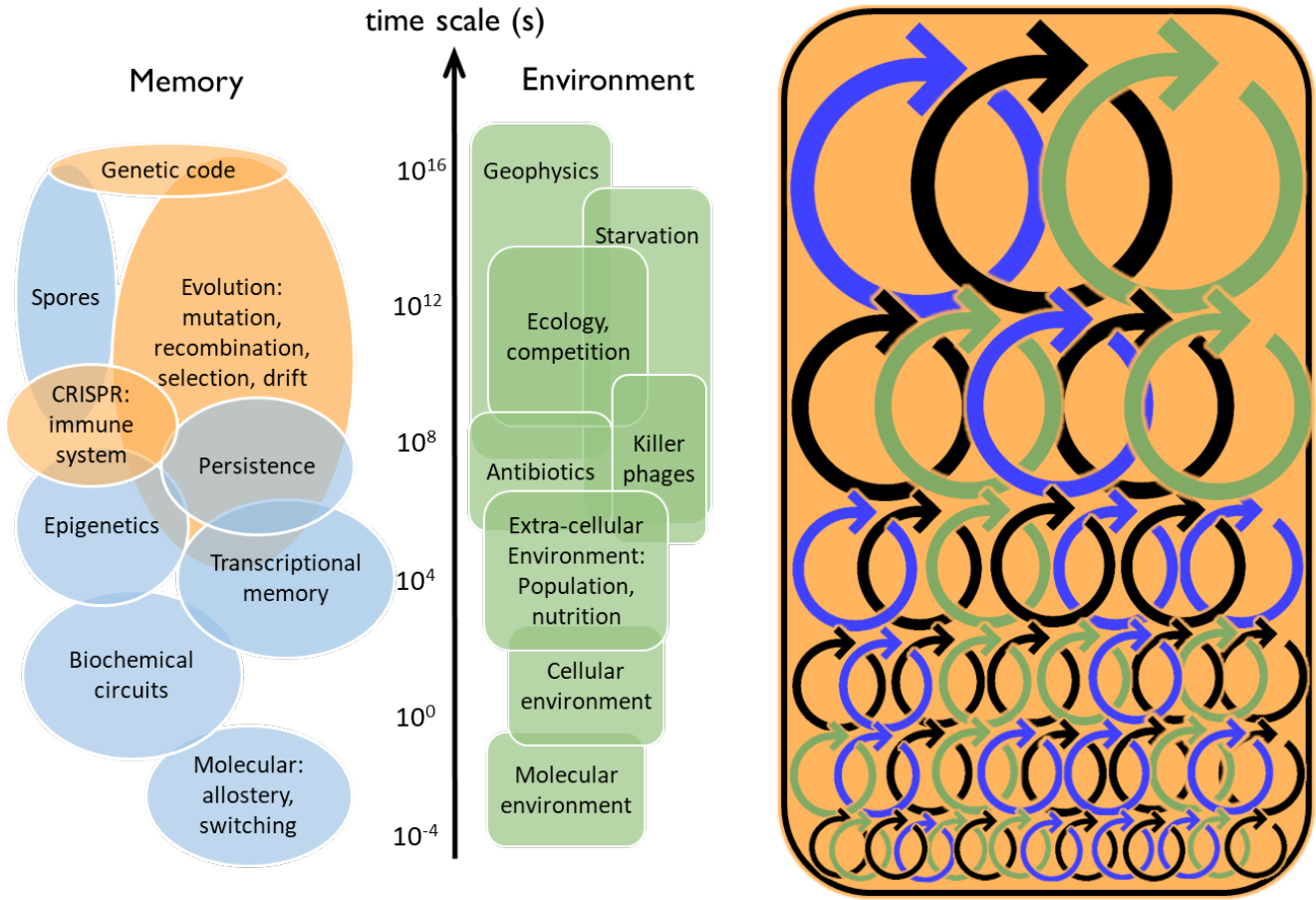

**Figure 2. The memory cascade of *E. coli*. Right**: An artistic depiction of the hierarchy of memories, from short time scales (bottom) to longer ones (top). The loops represent the back-and-forth feedback between the memory levels. The size of the loop represents the timescale (logarithmically). **Left:** the spread of memory time-scales, from the shortest biochemical scales (milliseconds) to the longest evolutionary time-scale of billions of years (1 year ~ $3 \cdot 10^7$ seconds). The vertical size of each ellipse (blue – phenotypic memory, orange – genetic) is roughly the timescale range of the corresponding memory. The green rectangles represent the environmental stresses mirrored in the memories.

As in turbulence, the faster timescales are the more dissipative and noisy ones. For example, the production and degradation of proteins in the cell is an error-prone process, strongly affected by thermal noise (with an error rate ~$10^{-4}$). But the replication of DNA, a much slower memory that changes at evolutionary timescales, relies on error-correction mechanisms that keep the errors at a much lower rate (~$10^{-8}$). In that sense, one may argue that lower levels in the cascade are closer to the dissipative realm of non-living matter at the nanoscale, the regime of viscous low-Reynolds flow. Climbing up the cascade to slower levels, living matter departs from this physical dissipative world towards the practically error-free world of digital computing [10, 11]. DNA genes are closer in nature to the binary memory of solid-state devices.

In parallel to each memory level, environmental processes are operating at similar timescales. *The environment and the memory mirror each other* (Figure 2): the memory is an *internal representation* of information that proved useful for survival during evolution. The fast millisecond scales of molecular memories are also the typical rates of biochemical changes, such as variations in pH and ion concentration, in the close vicinity of the molecule (say a sphere of radius ~10 nm). At a cellular scale (~1 micron ~$10^3$ nm), the circuitry responses at timescales comparable to the time it takes for macromolecules to diffuse across the cell. Transcription and epigenetic memories record environmental changes at ecological length (~1-$10^3$ m) and time ($10^4 - 10^6$ sec) scales, at the level of colonies and competition for resources among coexisting populations. Genetic memories are even slower and reflect evolutionary processes of mutation, recombination, and selection.

As a rule of thumb, one can say that the more severe is the environment, the more drastic and long-term is the bacterial response. Phenotypic memories respond to relatively rapid changes in the environment. Genetic memories respond to severe detrimental threats such as the invasion of killer phages. Finally, in the extreme circumstances when nutrients are lacking, some bacteria can metamorphose into endosperms. This *almost infinite memory plasticity*, spanning over many orders of magnitude, from microseconds to millions of years, is a fundamental hallmark of life, reflecting the need to survive. The following is a partial list of some of the cascade's levels with no attempt to give a fully accurate picture.

- *Biochemical memories* (~$10^{-3}$ seconds). The fastest memories of the bacterium are encoded in molecular states. The molecular memories are being constantly written and erased by allostery, switching, phosphorylation and other types of conformational and biochemical transitions and modifications, with timescales that range broadly around $10^{-3}$ seconds. A principle common to many of the molecular memories is switching among multiple metastable conformational states. A classic example is the binding of an oxygen molecule to hemoglobin, inducing allosteric switching that increases the affinity to bind more oxygen [12]. But there are also molecular memories with much longer terms. An extreme example is prions

[13], misfolded proteins that can transform normally-folded proteins into misfolded ones [13], thereby transmitting the misfolded form. The "prion memory" may sustain for years (~$10^8$ seconds).

- *Physiological memories* (~1 second). The biochemical circuitry of the bacterium operates at a timescale of seconds. Signaling and regulatory networks process and convey information at this rate. Synthesis of a protein takes ~10 seconds and the whole *E. coli* take about ~$10^3$ seconds to self-reproduce (cell cycle). Still, other circuits are designed to be much slower, such as the circadian clock with its slow rhythm (1 day ~ $10^5$ seconds).

- *Transcriptional memories* (~$10^4$ seconds). In a changing environment, this class of *phenotypic memory* allows the bacteria to maintain its physiological state and carry it over many generations [14]. The memory is written by self-sustaining regulatory feedback loops, producing proteins whose lifetime is longer than the cell doubling time [15]. A classic example of this type of cellular memory is in Monod's experiment, when the bacterial response to changes in available nutrients, for example, from galactose to lactose, is sustained through generations [16]. Remembering the history of the environment and the phenotype may help the bacterium adapt the response algorithm of its regulatory circuit to extreme fluctuations.

- *Epigenetic memories* (~$10^5$ seconds). In its general sense, epigenetics refers to any mechanism that perpetuates gene expression profiles and physiological cell state across cell divisions without changes in DNA sequence [17-19]. Thus, epigenetics includes the phenotypic memories of the gene regulatory network, together with other mechanisms of non-genetic inheritance, such as the transmission of small RNAs. Besides the mRNA inventory, daughter cells also inherit from their mother cell noncoding RNAs, such as siRNAs, piRNAs, and longer noncoding RNAs, regulating and interfering with gene expression. Although the terms "epigenetic" and "phenotypic" have much overlap, we list epigenetic memory, such as DNA methylation and histone modifications, separately.

- *Persistence memory* (~$10^6$ seconds). This memory operates at the level of the *population* by inducing *phenotypic variation*: Under the influence of antibiotics, lethal environmental stress that kills most bacteria, there is a small subpopulation, say 1%, in a metabolically dormant state that can survive the extinction [20]. These dormant bacteria called *persisters* were discovered by Bigger during WWII in colonies of *Staphylococcus aureus* that survived penicillin [21]. The persister state is a phenotypic memory switch, induced by expressing a few proteins, which encodes the survival strategy of the population: persisters appear only after a threshold antibiotic level is crossed [22, 23], then their lifetime is proportional to the antibiotic level, and they eventually disappear once the antibiotic is absent [24].

- *CRISPR: the bacterial immune system* (~$10^6$ seconds). Even more lethal stress is brought by phages, viruses that invade the bacteria. Faced with this threat, the bacteria evolved an immune system called CRISPR/Cas (CRISPR is an acronym for clustered regularly interspaced short palindromic repeats, and Cas are the CRISPR associate proteins) [25-27]. This prokaryotic immune system relies on a specialized

*genomic memory*: The invading phages are chopped into short fragments. These are inserted into the bacterial genome between regularly spaced palindromic repeats in a designated region of the chromosome that serves as the immune system's memory. Then, with the help of the Cas proteins, CRISPR-RNAs are produced and are used to silence invading phages. The CRISPR/Cas memory is particularly striking because it is a mechanism of incorporating an environmental signal (the genome of invading phages) directly into the bacteria's genome and keeping it over many generations.

- *Evolutionary memory* ($\sim 10^5 - 10^{16}$ seconds). This memory is encoded in the genome and changes slowly, via Darwinian dynamics of mutation, recombination, and selection, and by neutral genetic drift. Evolution is bound to exhibit long-term memory since the high-dimensional space of the genotype is too huge to be comprehensively explored, even by billions of years of adaptation [28-32]. By aligning genetic sequences, this history of adaptation is exposed as a phylogenetic tree whose root is the last universal common ancestor (LUCA).

- *Spore memory* ($\sim 10^{15}$ seconds). Close to the top of the memory cascade, this ultra-long term type of memory is activated when nutrients become very scarce. Under such extreme stress, some bacteria species, like *Bacillus subtilis*, enter into a metamorphological differentiation process and transform into endospores, which can practically live forever; in this sense, they are a singularity in the memory cascade. Ferdinand Cohn discovered endospores in 1875, and characterized the process of *metamorphosis* into endospores, called sporulation, which took about 8 hours in the *Bacillus subtilis* he studied [33, 34]. Cohn used this striking discovery to fight the idea of spontaneous generation, together with his contemporaries Pasteur and Tyndall. Cohn showed that endospores could survive in hot water at $80^0$ C for days, and recent studies extended this timescale to millions of years [35, 36]. At the population level, the fraction of cells that sporulate is a short-term memory that depends on the physiological state of previous generations [37].

- *Frozen memories* ($>10^{16}$ seconds). These memories are so hard to change that they remain frozen. The classic example is the genetic code [38]: Indeed, the assignment of amino acids to codons exhibits patterns of adaptation for improved error-resilience [39]. But shortly after this early phase of evolution, before the last universal common ancestor (LUCA), the code froze and is therefore universal for almost all organisms on Earth, apart from a few dozen rare variations [40, 41].

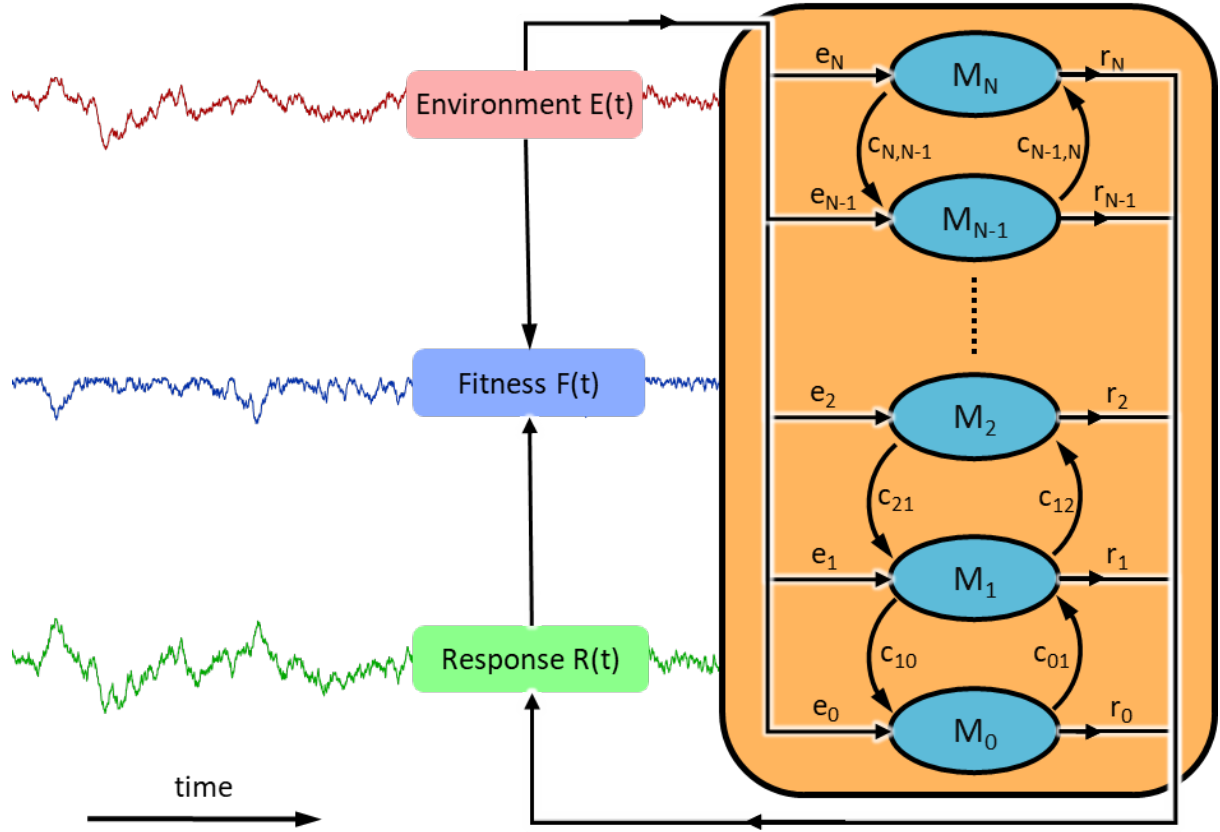

**Figure 3:** A toy model in the spirit of loop cascade (Figure 2). A cascade of $N+1$ memories, $M_0$ to $M_N$ with increasing time scopes, $\tau_0 \leq \tau_1 \leq \ldots \leq \tau_N$. The memories are interconnected by the coupling functions $c_{ij}$ that determine the feedback or feedforward from $M_i$ to $M_j$. The environment is represented by time series $E(t)$ which may affect any of the levels through the couplings $e_i$. The response $R(t)$ is computed from the states of the memories $M_i(t)$ through the output functions $r_i$. The survivability of the organism is determined by the fitness $F(t)$.

***Toy model: an evolving memory cascade.*** The image of an evolving memory cascade led us to construct a simple toy model where this idea can be examined more quantitatively. The goal is to mimic the generic features of the cascade. Thus, the only certain conceptual similarity is kept, and one should abandon the illusion that it simulates real living matter. Inspired by the chemotactic response example, we construct a battery of memories, from the shortest one $M_0$ to the longest one $M_N$ (see details in Figure 3). Each level $M_i$ is coupled to the levels above and below, $M_{i-1}$ and $M_{i+1}$, in a 1D linear cascade, but in general there could also be longer-range crosstalk with a higher effective dimension.

The environment is a discrete-time series $E(t)$, fluctuating in a range between $-1$ and $1$, which can feed into any of the memory levels (an equivalent continuous-time model can also be formulated). To survive, the cascade needs to compute its response for the next step, $R(t+1)$, using the values of all memory functions $M_i(t)$. The fitness $F(t+1)$ takes into account the *quality* of the response $R(t+1)$ with respect to the environment $E(t+1)$, and the *cost* of constructing the cascade. With the fitness function, we can envision a population of memory cascades competing in a changing environment and surviving according to the quality and cost of their computed response.

To make the cascade model more convenient to calculate, we consider a specific realization where each of the memories $M_i$ (where $0 \leq i \leq N$) changes according to a simple dynamic equation:

$$M_i(t+1) = \Psi_i \left[ \sum_{j \neq i} c_{ij} \int^t ds\, M_j(s) m_{ij}(t-s, \tau_{ij}) + e_i \int^t ds\, E(s) m_{ie}(t-s, \tau_{ie}) + n_i(t) \right]. \quad (2)$$

The dynamic function is a nonlinear mapping, $\Psi_i[x]$, which takes into account: (i) signals sent by memories $M_j$ into $M_i$, which depend on the coupling $c_{ij}$ and on the memory kernel functions $m_{ij}(t, \tau_{ij})$ with their memory scopes $\tau_{ij}$; (ii) signals from the environment $E(t)$ received through the coupling $e_i$ and the environmental memory function $m_{ie}(t, \tau_{ij})$; (iii) noise $n_i(t)$ in the memory itself and in the signals it receives. In the case of discrete dynamics, all integrals denote discrete summation. The response $R(t+1)$ is determined by the present state of the whole cascade $M_i(t+1)$ with the couplings $r_i$ and noise $n_r(t+1)$:

$$R(t+1) = \Psi_r \left[ \sum_i r_i M_i(t+1) + n_r(t+1) \right]. \quad (3)$$

The model makes the following simplifying assumptions:
(a) A linear cascade: memories interact only with memories below and above, $c_{ij} = 0$ if $|i - j| > 1$.
(b) All memory kernel functions $m_{ij}$ have the same shape, just with different scopes $\tau_{ij}$.
(c) Each memory $M_i$ may integrate over the lower memory $M_{i-1}$ via $m_{i,i-1}(t, \tau_i)$ with a scope $\tau_i$. The time scopes increase up the cascade, $\tau_0 \leq \tau_1 \leq \ldots \leq \tau_N$.
(d) Each memory $M_i$ may receive as feedback the present value of the memory above $M_{i+1}(t)$ (i.e. $m_{i,i+1}$ is a delta function).

(e) All dynamic functions are the same sigmoidal map $\Psi_i[x] = \Psi[x] = \tanh[x]$ or $\text{sign}[x]$.

(f) $\Psi[x]$ are functions of the sum the contributions from memories, environment and noise (Eq. (2)).

(g) The memory functions act as linear transforms, i.e. convolution of external or internal signals, $E(t)$ or $M_j(t)$, with the memory kernels.

(h) The response function $\Psi_r[x]$ is also a sigmoidal map, $\Psi_r[x] = \Psi[x]$, which depends on the weighted sum of the memories $M_i$ (Eq. (3)).

Thus, the cascade becomes a simple composition of linear transforms akin to FFT (if we use exponential memory functions), and a nonlinear sigmoidal mapping $\Psi$, which together are fast and easy to calculate.

The fitness $F(t+1)$ depends on how well the response $R(t+1)$ fits the present state of the environment fitness $E(t+1)$ and on the cost of constructing the cascade. For simplicity, we use the Gaussian fitness:

$$F(t+1) = \exp\left[-w_f\left[R(t+1) - E(t+1)\right]^2 - w_e \sum_i e_i^2 - w_c \sum_{i,j} c_{ij}^2 - w_r \sum_i r_i^2\right], \quad (4)$$

with the weights $w_f$ for the deviation between environment and response and $w_e$, $w_c$, $w_r$, for the cost of the corresponding components in the circuitry of the cascade. With the fitness, we can consider the evolution of a population of cascades, each specified by its set $\Lambda$ of couplings, memory timescales, and noise sources, $\Lambda = (\tau_i, e_i, c_{ij}, r_i, n_i, n_r)$. Our general plan is to implement standard recipes of evolutionary dynamics via simulation and simple models and study how the statistics of the environment $E(t)$ and the competition lead to the emergence of a cascade. In principle, one expects an interplay between the cost of constructing and maintaining memories and the benefit of encoding in these memories useful information about past conditions. We have made preliminary simulations to demonstrate the feasibility of the cascade model. However, our intention here is merely to present it as a conceptual Gedankenexperiment. A more detailed and quantitative study will be reported elsewhere.

**Discussion.** By construction, the memory cascade model embraces an "evolutionary ergodic principle" (or evolutionary Occam's razor): Every feasible mechanism not forbidden by hard physical constraints will eventually be explored by evolution. Thus, we allow for all possible directions of information flux in the cascade. This intentionally disregards a long history of biological theories that excluded specific directions of interactions in the cascade. A prominent example is the postulate of a sharp division between genotypes and phenotypes, which recast in molecular terms Weismann's barrier between germ and somatic cells [42]: Genotypes are inherited but isolated from direct effects of the environment, apart from random mutations, while phenotypes interact with the environment, but any change they acquire cannot be inherited. However, incorporating pieces of phages straight into the genome, the CRISPR machinery violates the hypothesis of genotype isolation, while inherited epigenetic changes and cell states show that acquired phenotype can be transmitted (Lamarckian inheritance [43]). In accord, the memory cascade allows for direct environmental input at any level, and there are no a priory isolated memories. Likewise, the cascade does not follow the central dogma that forbids information flow from proteins to genes [44] and allows in principle all feedforward and feedback directions.

The memory cascade reflects the dual nature of living matter. Each microbe is a biochemical computation and self-production machine, and each protein in the microbe is by itself a sophisticated submachine. But at the same time, microbes and their proteins are also books that tell a long history of crises through which they survived. For example, the microbial immune system is the annals of viral pathogens that the species encountered during its evolution. Living matter is so dense with information because every organism carries the history of its species and its competition and cooperation with other species.

**References**

[1] M. Proust, À la Recherche du Temps Perdu, Gaston Gallimard Paris, 1923.

[2] D. Bohm, Wholeness and the Implicate Order, Routledge Classics, London & New York, 2002.

[3] V. Bacot, S. Perrard, M. Labousse, Y. Couder, E. Fort, Multistable Free States of an Active Particle from a Coherent Memory Dynamics, Phys. Rev. Lett., 122 (2019) 104303.

[4] Y. Couder, S. Protière, E. Fort, A. Boudaoud, Walking and orbiting droplets, Nature, 437 (2005) 208-208.

[5] N.C. Keim, J.D. Paulsen, Z. Zeravcic, S. Sastry, S.R. Nagel, Memory formation in matter, Rev. Mod. Phys., 91 (2019) 035002.

[6] H.C. Berg, D.A. Brown, Chemotaxis in Escherichia coli analysed by three-dimensional tracking, Nature, 239 (1972) 500-504.

[7] J.E. Segall, S.M. Block, H.C. Berg, Temporal comparisons in bacterial chemotaxis, Proc. Natl. Acad. Sci. U.S.A., 83 (1986) 8987.

[8] A. Celani, M. Vergassola, Bacterial strategies for chemotaxis response, Proceedings of the National Academy of Sciences, 107 (2010) 1391.

[9] U. Frisch, Turbulence : the legacy of A.N. Kolmogorov, Cambridge University Press, Cambridge, Eng. ; New York, 1995.

[10] A. Condon, H. Kirchner, D. Larivière, W. Marshall, V. Noireaux, T. Tlusty, E. Fourmentin, Will biologists become computer scientists?, EMBO reports, 19 (2018) e46628.

[11] R. Bar-Ziv, T. Tlusty, A. Libchaber, Protein–DNA computation by stochastic assembly cascade, Proc. Natl. Acad. Sci. U.S.A., 99 (2002) 11589-11592.

[12] J. Monod, J.-P. Changeux, F. Jacob, Allosteric proteins and cellular control systems, J. Mol. Biol., 6 (1963) 306-329.

[13] D.C. Gajdusek, C.J. Gibbs, M. Alpers, Experimental Transmission of a Kuru-like Syndrome to Chimpanzees, Nature, 209 (1966) 794-796.

[14] E. Jablonka, B. Oborny, I. Molnar, E. Kisdi, J. Hofbauer, T. Czaran, The adaptive advantage of phenotypic memory in changing environments, Philosophical Transactions of the Royal Society of London. Series B: Biological Sciences, 350 (1995) 133-141.

[15] G. Lambert, E. Kussell, Memory and fitness optimization of bacteria under fluctuating environments, PLoS Genet., 10 (2014).

[16] J. Monod, The growth of bacterial cultures, Annu. Rev. Microbiol., 3 (1949) 371-394.

[17] E. Heard, Robert A. Martienssen, Transgenerational Epigenetic Inheritance: Myths and Mechanisms, Cell, 157 (2014) 95-109.

[18] E.J. Richards, Inherited epigenetic variation — revisiting soft inheritance, Nature Reviews Genetics, 7 (2006) 395-401.

[19] J. Casadesús, R. D'Ari, Memory in bacteria and phage, Bioessays, 24 (2002) 512-518.

[20] E. Maisonneuve, K. Gerdes, Molecular Mechanisms Underlying Bacterial Persisters, Cell, 157 (2014) 539-548.

[21] J. Bigger, Treatment of staphylococcal infections with penicillin, Lancet, 2 (1944) 497-500.

[22] N. Balaban, J. Merrin, R. Chait, L. Kowalik, S. Leibler, Bacterial persistence as a phenotypic switch, Science, 305 (2004) 1622-1625.

[23] O. Gefen, N. Balaban, The importance of being persistent: heterogeneity of bacterial populations under antibiotic stress, FEMS Microbiol. Rev., 33 (2009) 704-717.

[24] E. Rotem, A. Loinger, I. Ronin, I. Levin-Reisman, C. Gabay, N. Shoresh, O. Biham, N. Balaban, Regulation of phenotypic variability by a threshold-based mechanism underlies bacterial persistence, Proc. Natl. Acad. Sci. U.S.A., 107 (2010) 12541-12546.

[25] F. Mojica, C. Diez-Villasenor, J. Garcia-Martinez, E. Soria, Intervening sequences of regularly spaced prokaryotic repeats derive from foreign genetic elements, J. Mol. Evol., 60 (2005) 174-182.

[26] A. Bolotin, B. Ouinquis, A. Sorokin, S. Ehrlich, Clustered regularly interspaced short palindrome repeats (CRISPRs) have spacers of extrachromosomal origin, Microbiology, 151 (2005) 2551-2561.

[27] R. Barrangou, L. Marraffini, CRISPR-Cas systems: prokaryotes upgrade to adaptive immunity, Mol. Cell, 54 (2014) 234-244.

[28] I.S. Povolotskaya, F.A. Kondrashov, Sequence space and the ongoing expansion of the protein universe, Nature, 465 (2010) 922-926.

[29] T. Tlusty, A. Libchaber, J.-P. Eckmann, Physical Model of the Genotype-to-Phenotype Map of Proteins, Physical Review X, 7 (2017) 021037.

[30] T. Tlusty, Self-referring DNA and protein: A remark on physical and geometrical aspects, Philosophical Transactions of the Royal Society A: Mathematical, Physical and Engineering Sciences, 374 (2016) 20150070.

[31] S. Dutta, J.-P. Eckmann, A. Libchaber, T. Tlusty, Green function of correlated genes in a minimal mechanical model of protein evolution, Proc. Natl. Acad. Sci. U.S.A., 115 (2018) E4559-E4568.

[32] J.-P. Eckmann, J. Rougemont, T. Tlusty, Colloquium: Proteins: The physics of amorphous evolving matter, Rev. Mod. Phys., 91 (2019) 031001.

[33] F. Cohn, Untersuchungen nuch bacterium, Beiträge zur Biologie der Pflanzen, 3 (1875) 141-207.

[34] F. Cohn, Die Pflanze. Voträge aus dem Gebiete der Botanik, J. U. Kern (M. Müller), Breslau,, 1896.

[35] R.H. Vreeland, W.D. Rosenzweig, D.W. Powers, Isolation of a 250 million-year-old halotolerant bacterium from a primary salt crystal, Nature, 407 (2000) 897-900.

[36] R. Cano, M. Borucki, Revival and identification of bacterial spores in 25- to 40-million-year-old Dominican amber, Science, 268 (1995) 1060-1064.

[37] J.-W. Veening, E.J. Stewart, T.W. Berngruber, F. Taddei, O.P. Kuipers, L.W. Hamoen, Bet-hedging and epigenetic inheritance in bacterial cell development, Proc. Natl. Acad. Sci. U.S.A., 105 (2008) 4393-4398.

[38] F.H.C. Crick, The origin of the genetic code, J. Mol. Biol., 38 (1968) 367-379.

[39] T. Tlusty, A colorful origin for the genetic code: Information theory, statistical mechanics and the emergence of molecular codes, Phys Life Rev, 7 (2010) 362-376.

[40] S. Osawa, T.H. Jukes, K. Watanabe, A. Muto, Recent evidence for evolution of the genetic code, Microbiological Reviews, 56 (1992) 229-264.

[41] B.G. Barrell, A.T. Bankier, J. Drouin, A different genetic code in human mitochondria, Nature, 282 (1979) 189-194.

[42] A. Weismann, Das Keimplasma: eine theorie der Vererbung, Fischer1892.

[43] J.B.P.A.d.M.d. Lamarck, Zoological philosophy : an exposition with regard to the natural history of animals, University of Chicago Press, Chicago, 1984.

[44] F.H. Crick, On protein synthesis, Symp Soc Exp Biol, 1958, pp. 138-163.